\documentclass{amsart}
\usepackage{amsmath}
\usepackage{latexsym}
\usepackage{amsfonts}

\usepackage{amssymb}
\usepackage{amscd}
\usepackage{amsgen,amstext,amsbsy,amsopn}

\usepackage{pdfsync}

\newtheorem{lemma}{LEMMA}
\newtheorem{theorem}{THEOREM}

\begin{document}

\title{DISORDERED BOSE EINSTEIN CONDENSATES\\ WITH INTERACTION}
\thanks{Contribution to the proceedings of ICMP12, Aalborg, Denmark, August 6--11, 2012. \\ \copyright\, 2012 by
  the authors. This paper may be reproduced, in its entirety, for
  non-commercial purposes.}

\author{ROBERT SEIRINGER}
 \address{Department of Mathematics and Statistics, McGill University, \\ 805 Sherbrooke Street West, Montreal, QC H3A 2K6, Canada\\
E-mail: robert.seiringer@mcgill.ca}

\author{JAKOB YNGVASON}
 \address{Fakult\"at f\"ur Physik, Universit{\"a}t Wien,	\\ Boltzmanngasse 5, 1090 Vienna, Austria\\
E-mail: jakob.yngvason@univie.ac.at}

\author{VALENTIN A.\ ZAGREBNOV}\address{D\'{e}partement de Math\'{e}matiques,
Universit\'e d'Aix-Marseille (AMU) and Centre de Physique
Th\'eorique - UMR 7332,  Luminy Case 907, 13288 Marseille, Cedex09, France\\
E-mail: valentin.zagrebnov@univ-amu.fr}

\begin{abstract}
  We study the effects of random scatterers on the ground state of the
  one-dimensional Lieb-Liniger model of interacting bosons on the unit
  interval in the Gross-Pitaevskii regime. We prove that Bose Einstein
  condensation survives even a strong random potential with a high
  density of scatterers. The character of the wave function of the
  condensate, however, depends in an essential way on the interplay
  between randomness and the strength of the two-body interaction. For
  low density of scatterers or strong interactions the wave function
  extends over the whole interval. High density of scatterers and weak
  interaction, on the other hand, leads to localization of the wave
  function in a fragmented subset of the interval.

\end{abstract}

\maketitle

%%%%%%%%%%%%%%%%%%%%%%%%%%%%%%%%%%%
% \newcommand{\half}{\mbox{$\frac{1}{2}$}}
% \newcommand{\begin{equation}}{\begin{equation}}
%     \newcommand{\end{equation}}{\end{equation}}
%%%%%%%%%%%%%%%%%%%%%%%%%
\section{Introduction}
While the effects of random potentials on single particle
Schr\"odinger operators \cite{pasturfigotin} and ideal Bose
gases \cite{lenoblepasturzagrebnov,LZ,JPZ}
% \cite{lenoblepasturzagrebnov}--\cite{JPZ}
are rather well explored, the present understanding of such effects on
many-body systems of interacting particles is much less complete.  In
recent years, however, many papers concerning the interplay of
Bose-Einstein Condensation (BEC) and disorder have appeared of which
references [\cite{gimperlein}--\cite{BW}] are but a
sample.

We present here results on a model that is in a sense the simplest one
imaginable where this interplay can be studied by rigorous
mathematical means. This is the one-dimensional Lieb-Liniger (LL)
model \cite{LL} of bosons with contact interaction in a `flat' trap,
augmented by an external random potential that is generated by Poisson
distributed point scatterers of equal strength.  We study the ground
state and prove that, no matter the strength of the random potential,
BEC is not destroyed by the random potential in the Gross-Pitaevskii
(GP) limit where the particle number tends to infinity while the
coupling parameter in a mean-field scaling stays fixed.  The character
of the wave function of the condensate, however, depends in an
essential way on the relative size of the three parameters
involved. These are the scaled coupling parameter $\gamma$ for the
interaction among the particles, the density of the scatterers $\nu$,
and the strength $\sigma$ of the scattering potential. All these
parameters are assumed to be large in suitable units.

Our main results an be summarized as follows: For $\sigma\gg 1$ we
identify three different ``phases" of the condensate. For large
$\gamma\gg \nu^2$ the condensate is extended over the whole trap (the
unit interval in our model). A transition from a delocalized to a
localized state takes place when $\gamma$ is of the order $\nu^2$, in
the sense that for $\gamma\ll \nu^2$ the density is essentially
distributed among a fraction $\lambda \ll 1$ of the $\nu\gg 1$
intervals between the obstacles. For $\gamma\gg\nu/(\ln \nu)^2$ we
still have $\lambda\nu\gg 1$, but for $\gamma\sim\nu/(\ln \nu)^2$ the
fraction of intervals that are significant occupied shrinks to
$O(\nu^{-1})$.\footnote{The notation $a\sim b$ means that $a/b$ tends to a finite, nonzero constant.} We stress, however, that in all cases there is complete
BEC into a single state in the limit when the particle number tends to
infinity.

Details of our analysis are given in the paper \cite{SYZ}.

\section{The Model}

The model we consider is the {Lieb-Liniger model} of bosons with
contact interaction on the unit interval but with an additional
{external random potential} $V_\omega$. The Hamiltonian on the Hilbert
space $L^2([0,1], dz)^{\otimes_{\rm s}N}$ is
\begin{equation} \label{model}
  H=\sum_{i=1}^N\left(-\partial_{z_i}^2+V_\omega(z_i)\right)+\frac{{\gamma}}
  N\sum_{i<j}\delta(z_i-z_j)
\end{equation}
with $\gamma\geq 0$ and Dirichlet boundary conditions at the end points of the interval. The random
potential is taken to be
\begin{equation}
  V_\omega(z)=\sigma
  \sum_j\delta(z-z_j^\omega)
\end{equation}
with $\sigma \geq 0$ independent of the random sample $\omega$ while
the \emph{obstacles} $\{z_j^\omega\}$ are \emph{Poisson distributed}
with density $\nu\gg 1$, i.e., their mean distance is $\nu^{-1}$. 

The Hamiltonian \eqref{model} can be defined rigorously via the quadratic form on the Sobolev space  $H^1([0,1]^{\times N})$ given by the expression on the right hand side of \eqref{model}, noting that functions in the Sobolev space can be restricted to hyperplanes of codimension
1. See also \cite{BEH}, Ch. 14.6. The limiting case $\sigma=\infty$ amounts to requiring
the wave function to vanish at the positions of the obstacles $z^\omega_j$.

% ========================================================
There are several reasons for studying this model:
\begin{itemize}
\item It is the simplest model of its kind.
  % \item The 1D LL model can be obtained as a limit of a 3D model
  %   (LSY, 2003; Seiringer, Yin, 2008) and is also experimentally
  %   realizable.
\item BEC in the ground state can be proved in a suitable limit, for
  an \emph{arbitrary} nonnegative external potential.
\item $V_\omega$ is simple enough to allow a rigorous analysis of the
  condensate.
\end{itemize}
We remark also that the case $\gamma=0$, $\sigma=\infty$ corresponds
to the \emph{Luttinger-Sy model} \cite{LuSy}.

Once BEC has been established the main question concerns the
dependence of the properties of the condensate on the three parameters
$\gamma$, $\sigma$ and $\nu$.

% ========================================================

\section{BEC in the Ground State}

A basic nontrivial fact about the model \eqref{model}, whose proof will be discussed below, is that for \emph{fixed}
$\gamma$, $\sigma$ and configuration $\omega$ there is \emph{complete
  BEC in the ground state} in the sense that the 1-particle density
matrix/$N$ converges to a one dimensional projector as
$N\to\infty$. As usual, the 1-particle density matrix of the ground state wave function $\Psi_0$ is defined as
\begin{equation}
\gamma_0(z,z')=N\int\Psi_0(z,z_2,\dots,z_N)\overline{\Psi_0(z,z_2,\dots,z_N)}\,dz_2\cdots dz_N.
\end{equation}
Furthermore, the corresponding wave function of the condensate, i.e., the eigenfunction to the highest eigenvalue $O(N)$ of the integral operator defined by $\gamma(z,z')$ is, in the limit $N\to\infty$,  the
$L^2$-normalized minimizer of the \emph{Gross-Pitaevskii (GP) energy
  functional}
\begin{equation}\label{GPfunct}
  \mathcal E^{\rm
    GP}[\psi]=\int_0^1\left\{|\psi'(z)|^2+V_\omega(z)|\psi(z)|^2+(\gamma/2)|\psi(z)|^4\right\}dz.
 \end{equation}
 Formally, \eqref{GPfunct} is obtained by computing the expectation value of $H/N$ with a Hartree-type wave function $\Psi=\psi^{\otimes N}$ and taking the limit $N\to\infty$.
 
Since we want to consider large values of $\nu$, $\sigma$ and $\gamma$
 it is important to estimate also the {rate of the
  convergence} of the 1-particle density matrix as $N\to\infty$, in
dependence of these parameters and of the configurations $\omega$.

Our proof of BEC in the GP limit is simpler than the corresponding proof in three or
two dimensions \cite{LS} because the one-dimensional case considered here corresponds to a
{\it high density}, mean-field limit. In contrast, the work in \cite{LS} deals with a {\it low density} limit that
requires quite different tools.

% ========================================================

\subsection{The Proof of BEC (sketch)}

The proof of BEC is based on energy bounds:
\begin{itemize}
\item An {upper bound} to the many-body {ground state energy}
  $E_0^{\rm QM}$ by taking $\psi_{0}^{\otimes N}$ as a trial function
  for $H$ where $\psi_{0}$ is the minimizer of the GP energy
  functional, normalized so that $\Vert \psi_0\Vert_2=1$. This gives
  \begin{equation}
    E_0^{\rm QM}\leq N e_0
    % (1+o(1))
  \end{equation}
  where $e_0=\mathcal E^{\rm GP}[\psi_{0}]$ is the g.s.e.\ of the GP
  functional.
  % This estimate is uniform in the external potential $V_\omega\geq
  % 0$.

\item An \emph{operator lower bound} for the many-body Hamiltonian $H$,
  up to controlled errors, in terms of the 1-particle {mean-field
    Hamiltonian}
  \begin{equation}\label{meanfieldham}
    h=-\partial_z^2+V_\omega(z)+\gamma|\psi_0(z)|^2-(\gamma/2)\hbox{$\int$}
    |\psi_0|^4
  \end{equation}
  which has $\psi_0$ as ground state with energy
  $e_0$. For this bound similar ideas
as in the proof of Proposition 6.4 in \cite{BSY} are used.%This bound is the tricky part!
\end{itemize}
% The energy bounds are uniform in $V_\omega$.

% ========================================================
BEC follows from the upper and lower bounds and the fact that there is
an {energy gap} between $e_0$ and the next lowest eigenvalue, $e_1$,
of the mean-field Hamiltonian $h$:

Let
\begin{equation}
  N_0=\int \overline{\psi_0(z)}\gamma_0(z,z')\psi_0(z')\,dzdz'
 =\langle \Psi_0|
 a^\dagger(\psi_0)a(\psi_0)|\Psi_0\rangle
\end{equation}
be the occupation number of the GP ground state $\psi_0$ in the
many-body ground state $\Psi_0$. Then the energy bounds give
\begin{equation}
  N_0e_0+(N-N_0)e_1-o(1)Ne_0 \leq E_0^{\rm QM}\leq
  Ne_0\,
\end{equation}
where the $o(1)$ factor depends only on $\gamma$ and $N$.  This
implies an {upper bound for the depletion}:
\begin{equation}
  \left(1-\frac {N_0}N\right)\leq
  {o(1)}\frac{e_0}{e_1-e_0}
\end{equation}
% The right-hand side, {averaged} over $\omega$, can be shown to tend
% to $0$ if $N\to\infty$ and the parameters $\gamma,\nu,\sigma$ do not
% grow too fast with $N$.

% ========================================================
% ========================================================

More precisely, the estimate  on the depletion of the condensate proved in \cite{SYZ} is

\begin{theorem}[BEC]
  \begin{equation} \label{depletion} \left(1-\frac {N_0}N\right)\leq C
    \frac {e_0}{e_1-e_0} N^{-1/3}\min\{\gamma^{1/2},\gamma\}.
  \end{equation} 
\end{theorem}

We must, however, also consider the dependence of the {energy gap},
$e_1-e_0$, on the random potential and the parameters.
\subsection{Remark} If $e_0<e_1\leq \cdots\leq e_k$ are the $(k+1)$ lowest eigenvalues of \eqref{meanfieldham} with corresponding eigenfunctions $\psi_j$ and
\begin{equation}N_{< k}=\sum_{j=0}^{k-1}\langle \Psi_0|
 a^\dagger(\psi_j)a(\psi_j)|\Psi_0\rangle\end{equation}
is the occupation of the $k$
lowest eigenstates with energies $\leq e_{k-1}$,
then we have
\begin{equation} \label{kdepletion} \left(1-\frac {N_{<
        k}}N\right)\leq C \frac {e_0}{e_k-e_0}
  N^{-1/3}\min\{\gamma^{1/2},\gamma\}.
\end{equation} 
For finite $N$ this estimate may be more useful than \eqref{depletion}
because even if $k\ll N$, $e_k-e_0$ can be substantially larger than
$e_1-e_0$. Thus the right side of \eqref{kdepletion} can be $<1$ even
in cases when \eqref{depletion} contains no information because the
right-hand side is $>1$.

A situation when $1-N_{<k}/N$ is small for some $1\ll k\ll N$ without 
$1-N_{0}/N$ being small is commonly referred to as a fragmented condensation. For finite $N$
this may be a reasonable substitute for the full condensation
that in our model emerges in the large $N$ limit according to Theorem 3.1.  We also point out that as
far as the particle density $\rho(z)=\gamma(z,z)$ in position space is concerned, a fragmented condensate with
non-overlapping single-particle wavefunctions is indistinguishable from a fully condensed
state where the wavefunction of the condensate is a coherent superposition of the nonoverlapping
functions. The difference shows up, however, in the density
$\hat\rho(p)=(2\pi)^{-1}\int\exp(\mathrm ip(z-z'))\gamma(z,z')\,dzdz'$ in momentum space.

% ========================================================

\subsection{The energy gap}
Consider a one-dimensional Schr\"odinger operator $-\partial_z^2+W(z)$
on the unit interval with a nonnegative potential $W$ and Dirichlet
boundary conditions.

\begin{lemma}[Gap]
  Define $\eta>0$ by $\eta^2 = \pi^2 + 3 \int_0^1 W(z) dz.$ Then
  \begin{equation}
    e_1 - e_0 \geq \eta\, \ln \left( 1 + \pi
      e^{-2\eta}\right)
  \end{equation}
\end{lemma}
 
The proof is based on a modification of a result of Kirsch and Simon
\cite{kirschsimon}, that involves the sup norm of $W$ instead of the
integral.  In our case
$\eta=\eta_\omega=\sqrt{\pi^2+3m_\omega\sigma+3\gamma}$ where
$m_\omega$ is the number of obstacles in $[0,1]$, that is almost
surely equal to $\nu$ in the limit $\nu\to\infty$.

For large $\sigma m_\omega $ the bound is certainly not optimal, in
fact, in this case one expects $e_1 - e_0\sim (\sigma m_\omega)^{-1}$.

% ========================================================
%

%
% ========================================================

\subsection{The Poisson distribution of the obstacles}

Let $z_1^\omega\leq z_2^\omega\leq \dots \leq z_{m_\omega}^\omega$
denote those random points which lie in the unit interval $[0,1]$.
The lengths $\ell_i=z^{\omega}_{i+1}-z^{\omega}_i$ are independent
random variables with distribution
\begin{equation}
  dP_\nu(\ell) = \nu e^{-\ell\nu} d\ell
\end{equation}
and we are considering the case $\nu\gg 1$.

The average length of an interval free of obstacles is $\nu^{-1}$ and
with probability one, $m_\omega/\nu\to 1$ and $\sum_i\ell_i=1$ for
$\nu\gg 1$. Combined with the estimate on the energy gap this implies
that the depletion of the BEC is {uniform} in the $L^p$ norm on sample
space for any $p<\infty$.
% if $\nu$, $\gamma$ and $\sigma$ are given.

\subsection{Ideal vs interacting gas}

While the average length of an interval is $\nu^{-1}$ there is, with
probability one, a unique largest interval of length $\sim \nu^{-1}
\ln \nu$. If there is no interaction, i.e., $\gamma=0$, and
$\sigma\to\infty$ (Luttinger-Sy model), then the ground state wave
function will be {localized} in the largest interval for kinetic
energy reasons, with energy $\sim \nu^2/(\ln \nu)^2$.

The question is how the character of the wave function of the
condensate, $\psi_0$, changes when the {interaction}, i.e., the term
$(\gamma/N)\int|\psi_0(z)|^4dz$ comes into play.

% ========================================================
\section{The Mass Distribution of the Condensate}
\subsection{A limit theorem for the GP energy}
Our first result is that the energy becomes \emph{deterministic} in an
appropriate limit.

Let $e_\omega(\gamma,\sigma,\nu)$ denote the GP energy with the random
potential $V_\omega$ and $e_0(\gamma,\nu)$ the energy for
$\sigma=\infty$\footnote{Recall that putting  $\sigma=\infty$ amounts to requiring the wave functions to vanish at the positions of
the obstacles.}, {averaged} over $\omega$.

\begin{theorem}[Convergence of the energy]
  Assume that $\nu\to \infty$, $\sigma\to \infty$ and $\gamma\to
  \infty$ in such a way that
  \begin{equation}\label{conditions}
    \gamma\gg \frac{\nu}{\left(\ln \nu\right)^2} \quad \text{\rm and} \quad 
    % \sigma \gg \min\{ \sqrt{\gamma\, f(\nu^2/\gamma)} , \ \nu\,
    % f(\nu^2/\gamma)\}\,.
    \sigma \gg \frac{ \nu}{1+ \ln \left( 1+ \nu^2/\gamma\right) }\,.
  \end{equation} 
  Then, for almost every sample $\omega$,
  \begin{equation}
    \lim \frac{
      e_\omega(\gamma,\sigma,\nu)}{e_0(\gamma,\nu)} = 1\,.
  \end{equation} 
\end{theorem}

\subsection{Comments on the proof}

The proof has two interrelated parts:
\begin{itemize}
\item Comparison of $e_\omega(\gamma,\sigma,\nu)$ with
  $e_\omega(\gamma,\infty,\nu)$.
\item Comparison of $e_\omega(\gamma,\infty,\nu)$ with the
  deterministic $e_0(\gamma,\nu)$.
\end{itemize}

Calculations are conveniently done in a grand canonical ensemble,
introducing a chemical potential $\mu$ that determines the optimal
repartition
\begin{equation}
  n(\ell)\approx(\ell\gamma)^{-1}[\mu\ell^2-\pi^2]_+
\end{equation}
of the condensate mass in intervals of length $\ell$ between the
obstacles. The lengths are distributed according to $dP_\nu(\ell)$ and
the normalization requires
\begin{equation}
  \nu\int n(\ell)dP_\nu(\ell)=1.
\end{equation}
% The constraint on $\sigma$ comes from the requirement that $\bar
% \ell\sigma$ is large where $\bar\ell$ is the average length of
% intervals for which $n(\ell)>0$.

% ========================================================

\subsection{Discussion of the GP wave function for large $\sigma$}

For large $\sigma$ the average number of intervals with non-zero
occupation numbers is given by
\begin{equation}
  \nu \int_{\pi/\sqrt\mu}^\infty dP_\nu(\ell) =
  e^{-\pi\nu/\sqrt\mu} \, \nu .
\end{equation}  
Since $\nu$ is the total (average) number of available intervals,
\begin{equation}
  \lambda :=e^{-\pi\nu/\sqrt\mu}\leq 1
\end{equation}
defines the {fraction} of them which are {occupied}. The normalization
requires
\begin{equation}\label{*}
  1 \sim \frac\mu\gamma e^{-\pi\nu/\sqrt\mu}\
\end{equation} 
so $\lambda$ is determined by $\gamma$ and $\nu$ via relation
\begin{equation}\label{**}
  \frac{\lambda }{\left(\ln\lambda^{-1}\right)^2} \sim \frac {\gamma}{ \nu^2}\,. 
\end{equation} 
We can now distinguish the following \emph{limiting cases\/}:
% ========================================================
\begin{itemize}
\item If $\gamma\gg \nu^2$ then by \eqref{**} we get $\lambda\to 1$,
  i.e., {\em all}\/ the intervals are occupied ({delocalization}). The
  chemical potential satisfies $\mu \sim \gamma$ in this regime.
\item If $\gamma\sim \nu^2$ then $\lambda \sim 1$, but $\lambda$ is
  strictly less than $1$ ({transition to localization}). Again we have
  $\mu \sim \gamma$.
\item If $\gamma\ll \nu^2$ then $\lambda \ll 1$, i.e., only a small
  fraction of the intervals are occupied ({localization}). The
  relation \eqref{*} implies $\mu \sim \frac{\gamma}\lambda$
  % \sim (\nu/\ln(\nu^2/\gamma))^2$
  for the chemical potential.
\item If $\gamma\sim \nu/(\ln\nu)^2$ then by \eqref{**} the fraction
  $\lambda$ becomes $O(1/\nu)$, i.e., {only finitely many intervals
    are occupied}. In this latter case, $\mu\sim \gamma\nu \sim
  \nu^2/(\ln\nu)^2$, which corresponds exactly to the inverse of the
  square of the size of the \textit{largest} interval.
\end{itemize}

% ========================================================

In particular, $\lambda \nu\gg 1$ only if $\gamma \gg \nu/(\ln
\nu)^2$, and hence this condition guarantees that many intervals are
occupied. In this case the {law of large numbers} applies and hence
the energy becomes {deterministic} in the limit.

If $\lambda \nu = O(1)$, on the other hand, the value of
$e_\omega(\gamma,\sigma,\nu)$ is {random}. This shows, in particular,
that our condition on $\gamma$ is optimal in the sense that for smaller $\gamma$ the
energy fluctuates.

Also the condition $\sigma\gg\nu/(1+\ln(1+\nu^2/\gamma))$ can be
expected to be optimal. It can be rephrased as $\bar \ell \sigma \gg
1$, where $\bar \ell$ is the (weighted) average length of occupied
intervals
\begin{equation}
  \bar \ell = \nu \int_0^\infty dP_\nu(\ell) \, \ell\, n(\ell)
\end{equation} 
with $n(\ell)$ the optimal repartition of the mass. A simple
calculation shows that $\bar \ell \sim \nu^{-1}
(1+\ln(1+\nu^2/\gamma))$.

\subsection{Comments on scaling and the thermodynamic limit}

Our model is formulated in the fixed interval $[0,1]$ so that the  particle density $\rho$ tends to infinity as $N\to\infty$. Equivalently, we could have considered the model in an interval $[-L/2, L/2]$ and taking $N$ and $L\to\infty$ with $\rho=N/L$, as done for instance in \cite{BW}. The two viewpoints are connected by simple scaling:

Let
\begin{equation} H_L=\sum_{i=1}^N\left(-\partial_i^2+ b\sum_j\delta (x_i-x_j^\omega)\right)+g\sum_{i<j}\delta(x_i-x_j)\end{equation}
be the Hamiltonian on the Hilbert space $L^2([-L/2, L/2], dx)^{\otimes_s N}$ with $x_j^\omega$ Poisson distributed with density $d$.
Writing $x_i=L z_i-(L/2)$ transforms $H_L$ into $L^{-2}H$ with $H$ the Hamiltonian \eqref{model} on $[0,1]$ and
\begin{equation}
\sigma=Lb,\quad \nu=Ld,\quad \gamma=LNg=L^2\rho g.
\end{equation}
The condition for BEC in Theorem 3.1 as well as the conditions 
\eqref{conditions} for the convergence of the GP energy to a deterministic value can straightforwardly be written as conditions for the parameters $b, d$ and $\rho g$ as $L\to\infty$. In particular, if $b$ and $d$ are fixed, then
\eqref{conditions} is fulfilled provided
\begin{equation} (L(\ln L)^2)^{-1}d\ll \rho g\ll d^2.\end{equation}
Note also that the validity of the GP approximation requires in any case that the dimensionless parameter $g/\rho$ is $\ll 1$, cf., e.g., \cite{LSY2004}. 

% ========================================================

\section{Conclusions}
\begin{itemize}
\item BEC in the ground state of the interacting gas in the GP regime
  can survive even in a strong random potential.
  % As far as BEC is concerned the interacting gas in this regime thus
  % behaves in a similar way as an ideal gas at zero temperature.
  The character of the wave function of the condensate, however, is
  strongly affected by the interaction.

\item A random potential may lead to {localization} of the wave
  function of the condensate in subintervals. The interparticle
  interaction counteracts this effect, however, and can lead to
  {complete delocalization} (the condensate extends over the whole
  unit interval) if the interaction is strong enough.

\item In terms of the interaction strength, $\gamma$, and density of
  scatterers, $\nu$, the {transition between localization and
    delocalization} occurs in the model considered when $\gamma\sim
  \nu^2$. For $\gamma\lesssim \nu/(\ln \nu)^2$ a {``third phase"}
  occurs where the condensate is localized in a small number of
  subintervals.
\end{itemize}

% \end{document}
%%%%%%%%%%%%%%%%%%%%%%%%%%%%%%%%%%

% \section{References}


\begin{thebibliography}{99}

\bibitem{pasturfigotin} L.A. Pastur, A. Figotin, \emph{Spectra of
    Random and Almost-Periodic Operators}, Springer-Verlag, 1992

\bibitem{lenoblepasturzagrebnov} O.\ Lenoble, L.A. Pastur,
  V.A. Zagrebnov, \emph{Bose-Einstein condensation in random
    potentials}, C.R. Physique \textbf{5}, 129--142 (2004)

\bibitem{LZ} O.Lenoble, V.A.Zagrebnov, \textit{Bose-Einstein
    Condensation in the Luttinger-Sy Model}, Markov Proc.Rel.Fields
  \textbf{13}, 441--468 (2007).

\bibitem{JPZ} Th.Jaeck, J.V.Pul\'{e}, V.A.Zagrebnov, \textit{On the
    Nature of Bose-Einstein Condensation Enhanced by Localization},
  J.Math.Phys. \textbf{51}, 103302-15 (2010).

\bibitem{gimperlein} H. Gimperlein, S. Wessel, J. Schmiedmayer,
  L. Santos, \emph{Ultracold Atoms in Optical lattices with Random
    One-Site Interactions}, Phys. Rev. Lett. \textbf{95}, 170401 (2005)

\bibitem{yukalovgraham} V.I. Yukalov, R. Graham, \emph{Bose-Einstein
    condensed systems in random potentials}, Phys. Rev. A \textbf{75},
  023619 (2007)

\bibitem{luganetal1} P. Lugan, P. Bouyer, A. Aspect, M. Lewenstein,
  L. Sanches-Palencia, \emph{Ultracold Bose Gases in 1D Disorder: From
    Lifshits Glass to Bose-Einstein Condensate}, Phys. Rev. Lett. {\bf
    98}, 170403 (2007)

\bibitem{fallanietal} L. Fallani, C. Fort, M. Inguscio,
  \emph{Bose-Einstein Condensates in Disordered Potentials},
  Adv. At. Molec. Opt. Phys. \textbf{56}, 119--159 (2008)

\bibitem{sanchesetal1} L. Sanches-Palencia, D. Cl\'ement, P. Lugan,
  P. Bouyer and A. Aspect, \emph{Disorder-induced trapping versus
    Anderson localization in Bose-Einstein condensates expanding in
    disordered potentials}, New J. Phys. \textbf{10}, 045019 (2008)

\bibitem{pikovskietal} A.S. Pikovsky, D.L. Shepelyansky,
  \emph{Destruction of Anderson Localization by a Weak Nonlinearity},
  Phys. Rev. Lett. \textbf{100}, 094101 (2008)

\bibitem{luganetal} P. Lugan, A. Aspect, L. Sanches-Palencia,
  D. Delande, B. Gr\'emaud, C.A. M\"uller, C. Miniatura,
  \emph{One-dimensional Anderson localization in certain correlated
    random potentials}, Phys. Rev. A \textbf{80}, 023605 (2009)

\bibitem{radicetal} J. Radic, V. Bacic, D. Judic, M. Segev, H. Buljan,
  \emph{Anderson localization of a Tonks-Girardeau gas in potentials
    with controlled disorder}, Phys. Rev. A \textbf{81}, 063639 (2010)

\bibitem{aleiner} I.L. Aleiner, B.L. Altshuler, G.V. Shlapnykov,
  \emph{A finite-temperature phase transition for disordered weakly
    interacting bosons in one dimension}, Nature Phys. \textbf{6},
  900--904 (2010)

\bibitem{sanchesetal2} L. Sanches-Palencia, M. Lewenstein,
  \emph{Disordered quantum gases under control}, Nature Phys., {\bf
    6}, 87--95 (2010)

\bibitem{piraudetal} M. Piraud, P. Lugan, B. Bouyer, A. Aspect, and
  L. Sanches-Palencia, \emph{Localization of a matter wave packet in a
    disordered potential}, Phys. Rev. A \textbf{83}, 031603(R) (2011)


\bibitem{cardosoetal} W.B. Cardoso, A.T. Avelar, D. Bazeia,
  \emph{Anderson localization of matter waves in chaotic potentials},
  Nonlin. Analysis \textbf{13}, 755--763 (2012)
  
\bibitem{SMBW} J. Stasinska, P. Massingnan, M. Bishop, J. Wehr, A. Sanpera and M. Lewenstein, \emph{The
glass to superfluid transition in dirty bosons on a lattice}, New J. Phys. {\bf 14},  043043 (2012)
  
\bibitem{BW} M. Bishop, J. Wehr,   \emph{Ground State Energy of Mean-field Model of Interacting Bosons
in Bernoulli Potential}, arXiv:1212.1487

 \bibitem{SYZ} R. Seiringer, J. Yngvason and V. A. Zagrebnov, \emph{Disordered Bose-Einstein condensates
with interaction in one dimension}, J. Stat. Mech. P11007 (2012), arXiv:1207.7054
  

\bibitem{LL} E.H. Lieb, W. Liniger, \emph{Exact Analysis of an
    Interacting Bose Gas. I.}, Phys. Rev. \textbf{130}, 1605--1616 (1963)
    
 

\bibitem{BEH} J. Blank, P. Exner, M. Havl\'{\i}\v{c}ek, Hilbert Space Operators in Quantum Physics,
Springer-Verlag, 2008

   \bibitem{LuSy} J.M.Luttinger, H.K.Sy,  \emph{Bose-Einstein Condensation in a One-Dimensional Model with Random Impurities},
Phys. Rev. A \textbf{7}, 712-720 (1973) 

\bibitem{LS} E.H. Lieb, R. Seiringer, \emph{Proof of Bose-Einstein
    Condensation for Dilute Trapped Gases}, Phys. Rev. Lett. \textbf{88},
  170409 (2002)

  \bibitem{BSY} B. Baumgartner, J.-P. Solovej and J. Yngvason, {\it
      Atoms in Strong Magnetic Fields: the High Field Limit at Fixed
      Nuclear Charge}, Commun. Math. Phys. \textbf{212}, 703--724
    (1998)

\bibitem{kirschsimon} W. Kirsch, B. Simon, \emph{Universal lower bounds
    on eigenvalue splittings for one dimensional Schr\"odinger
    operators}, Commun. Math. Phys. \textbf{97}, 453--460 (1985)

  % \bibitem{gronwall} P. Hartmann, \emph{Ordinary Differential
  %     Equations}, Wiley (1964)
  
\bibitem{LSY2004} E.H. Lieb, R. Seiringer and J. Yngvason,  {\it One Dimensional Behavior of
Dilute, Trapped Bose Gases},  Commun.Math.Phys. 
{\bf 244}, 347--393 (2004)


\end{thebibliography}
\end{document}